\documentclass[epjCONF,columns]{svjour}
\usepackage{graphics}
\usepackage[varg]{txfonts} 
\usepackage[latin1]{inputenc}
\usepackage{hyperref}
\usepackage{xspace}
\usepackage{subfig}
\usepackage{doi}

\providecommand {\TeV} {\ensuremath{\,}\textrm{Te\hspace{-.08em}V}\xspace}
\providecommand {\MeV} {\ensuremath{\,}\textrm{Me\hspace{-.08em}V}\xspace}
\providecommand {\GeV} {\ensuremath{\,}\textrm{Ge\hspace{-.08em}V}\xspace}

\providecommand{\PYTHIAeight} {{\textsc{pythia8}}\xspace}
\providecommand{\PYTHIAsix} {{\textsc{pythia6}}\xspace}

\providecommand{\PHOJET} {{\textsc{phojet}}\xspace}
\providecommand{\de}{\ensuremath{^\circ}}

\providecommand {\pt}       {\ensuremath{p_T}}

\providecommand {\deta}     {\ensuremath{\Delta\eta}}
\providecommand {\dphi}     {\ensuremath{\Delta\phi}}

\providecommand {\pp}    {\mbox{pp}}

\session-title{HCP2011, Paris, France}
\begin{document}
\title{Soft QCD from ATLAS and CMS}
\author{Romain Rougny\thanks{\email{romain.rougny@cern.ch}}, on behalf of ATLAS and CMS collaborations }
\institute{ Department of High-Energy Physics, Universiteit Antwerpen, Groenenborgerlaan 171, B-2020 Antwerpen, Belgium  }
\abstract{
Measurements of hadron production in pp collisions by the ATLAS and CMS experiments are presented, including charged particle transverse momentum, pseudorapidity and event-by-event multiplicity distributions at sqrt(s) = 0.9, 2.36 and 7 TeV, for NSD and inelastic events. Diffraction is studied with either diffraction enriched or suppressed data samples. Total inelastic cross-section as well as gap cross-section measurements are shown. Measured spectra of identified strange hadrons, reconstructed based on their decay topology, are also discussed. Comparisons to several QCD Monte Carlo models and tunes are exhibited. Results on two-particle angular correlations over a broad range of pseudorapidity and azimuthal angle in pp collisions are presented. Underlying event activity are studied with different hard probes: tracks, trackjets, calorimeter clusters, or in Drell-Yan events.
} 
\maketitle
\section{Introduction}\label{intro}
Since the beginning of the collisions at the LHC, the ATLAS and CMS experiments have measured many observables at $0.9, 2.36$ and $7\TeV$, through a vast physics program. Although the results presented here are mainly for soft $pp$ collisions, it is a mandatory step to better understand the physics involved, as QCD still needs to be modeled phenomenologically. As soft QCD is a background to many rare particle searches, improving our knowledge in this low-$\pt$ region for $7\TeV$ will be crucial to many analyses. Moreover, it is the first time data at such high energy has been acquired, thus allowing good tuning of the Monte Carlo generators. It also allows to discriminate between the handful of theoretical models that were built over the years, by allowing them to be confronted to real measurements.

\section{The ATLAS and CMS detectors}\label{detectors}
Complete descriptions of the ATLAS and CMS detectors can be found respectively in \cite{atlas:detector} and \cite{cms:detector}. Both are general-purpose detectors and were built for a broad range of analyses, from low-pt QCD physics to high mass Higgs searches. They are composed first of a silicon pixel and tracker detector covering an acceptance of $|\eta|<2.5$, plus a transition radiation tracker for ATLAS; then layers of electromagnetic and hadronic calorimeters, everything being embedded in a magnetic field of respectively 2 and 3.8 Tesla for ATLAS and CMS. The muon systems finish the onion-like subdetector layout of both experiments.

Minimum Bias events, where most of the inelastic cross-section is kept, were recorded with dedicated trigger detectors: the Minimum Bias Trigger Scintillator (MBTS, within $2.12<|\eta|<3.8$) for ATLAS, and a conjunction of the BPTX (located at $\pm175$m from the interaction point) and the Beam Scintillator Counters (BSC, within $3.23<|\eta|<4.65$) for CMS. In addition, beam background and beam halo events were discarded using triggers from those same detectors. For some CMS analyses where only non single diffractive (NSD) events were considered, the hadronic forward calorimeter (HF, within $2.9<|\eta|<5.2$) information was added to reject most of diffractive events. The results presented here use a wide range of integrated luminosity. The data samples from the $\pp$ collisions delivered by the LHC at the energies of $\sqrt{s}=0.9$ and $2.36\TeV$ during the years 2009-2010, and $\sqrt{s}=7\TeV$  during 2009 to 2011, were all analyzed.

\section{Results}\label{results}
\subsection{Minimum Bias results}\label{minbias}

\begin{figure}[Htbp]
\center{
\resizebox{1\columnwidth}{!}{
  \subfloat{\includegraphics{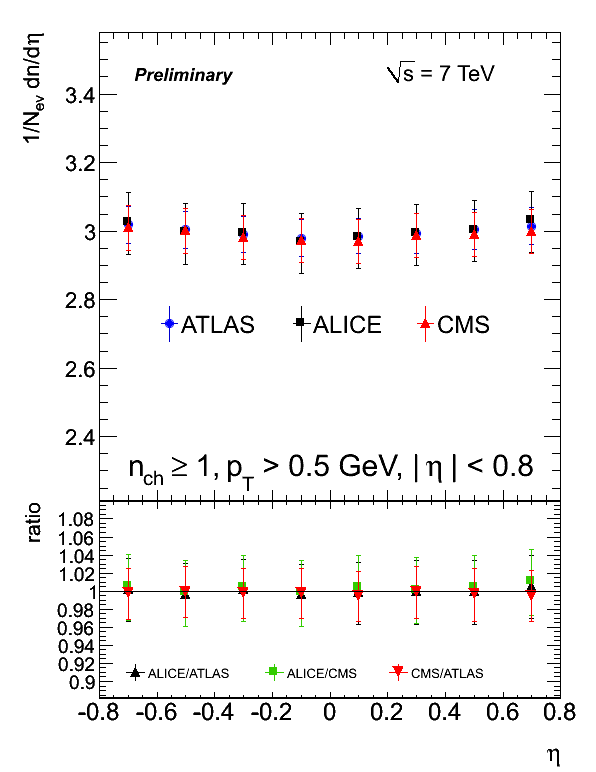}}
  \subfloat{\includegraphics{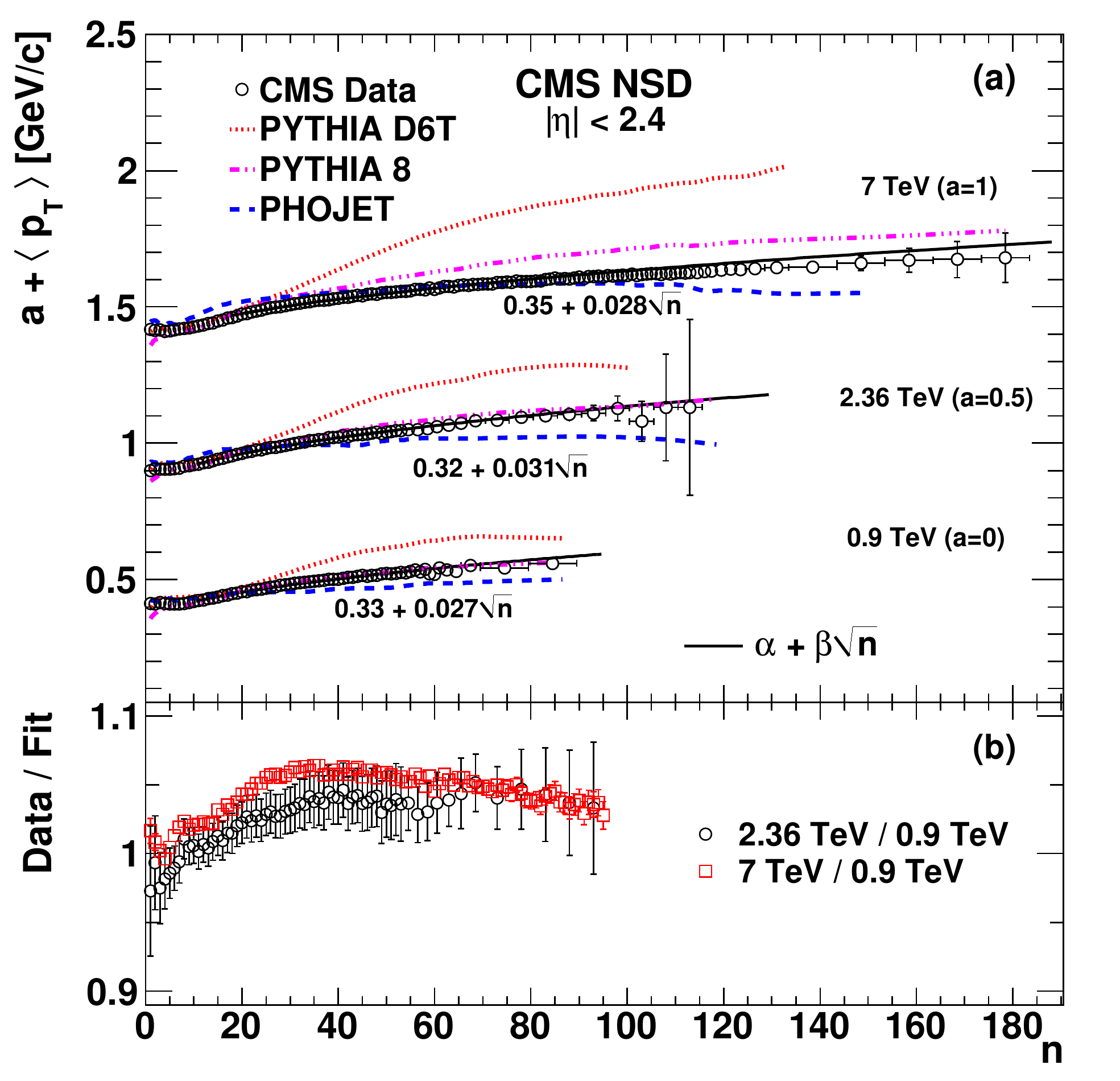}} }
\resizebox{1\columnwidth}{!}{
  \subfloat{\includegraphics{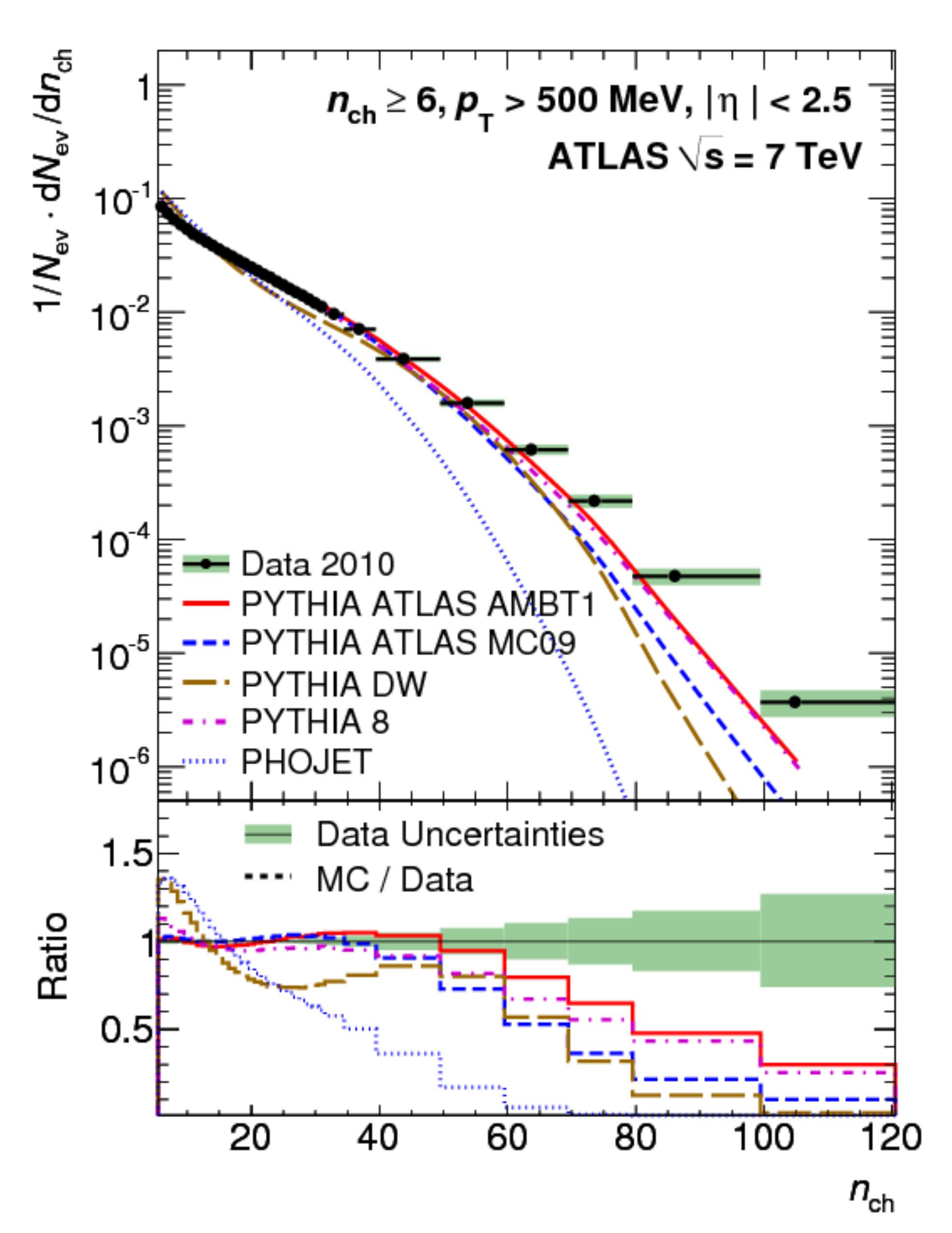}}
  \subfloat{\includegraphics{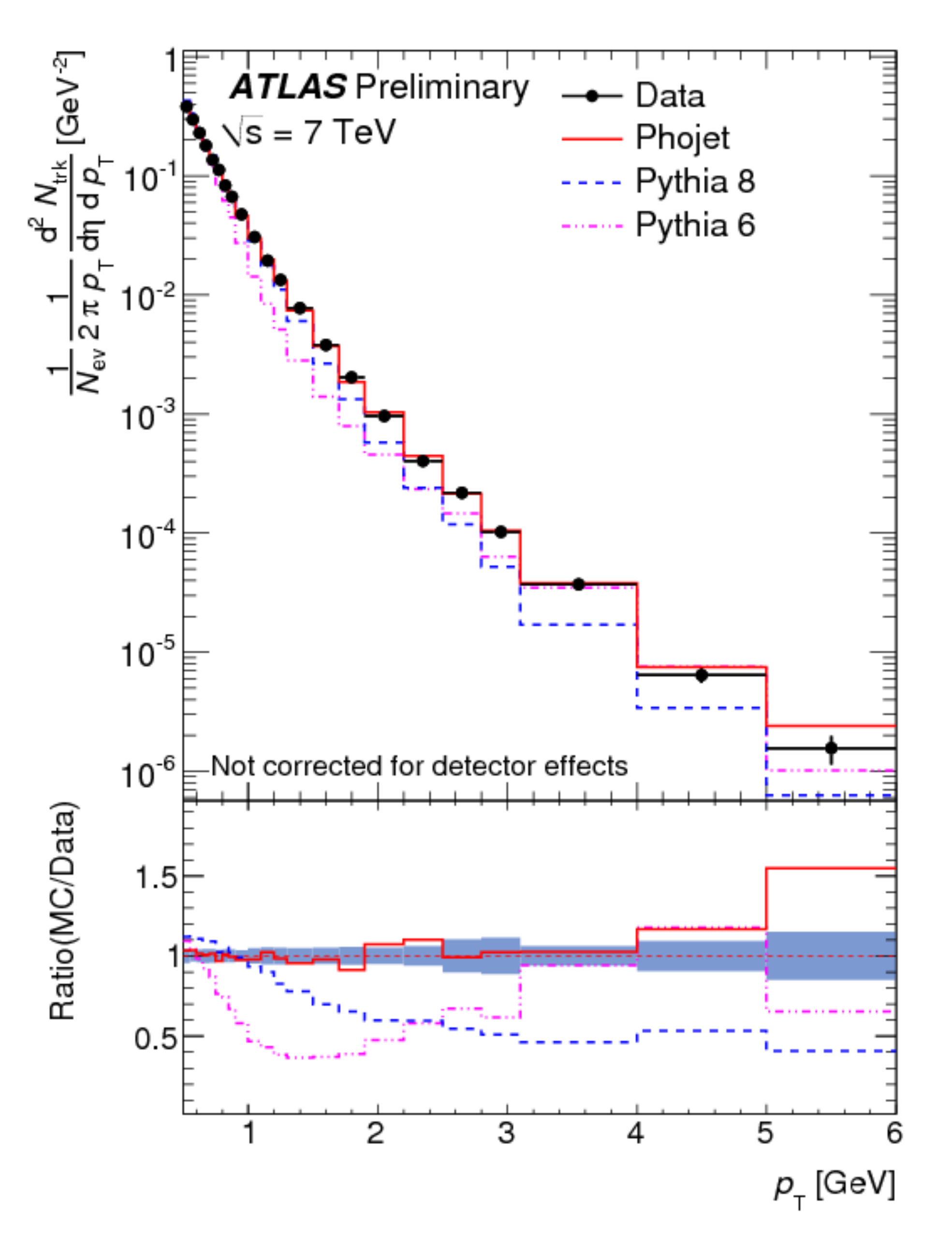}} }
}
\caption{\scriptsize{\bf(Top Left)} MBUEWG combination of the $\eta$ distribution at $7\TeV$ for ATLAS, ALICE and CMS, for events with at least one track of $\pt>0.5\GeV$ and $|\eta|<0.8$; {\bf(Top Right)} Evolution of the mean $\pt$ with the multiplicity for NSD events at 0.9, 2.36, and $7\TeV$; {\bf(Bottom Left)} Multiplicity distribution at $7\TeV$ for events with at least six tracks of $\pt>0.5\GeV$ and $|\eta|<2.5$, where diffraction should be greatly suppressed; {\bf(Bottom Right)} Uncorrected $\pt$ distribution at $7\TeV$ for diffraction enriched samples. Those minimum bias distributions are also compared with Monte Carlo predictions from different generators and tunes.}
\label{fig:mb}       
\end{figure}

Since it needed very few statistics, the first published observables from both experiments were the pseudorapidity ($\eta$) and transverse momentum ($\pt$) spectra\cite{atlas:dndeta_0.9tev,atlas:dndeta,cms:dndeta,cms:dndeta_7tev}. CMS chose an event selection rejecting diffractive events, and corrected up to NSD with Monte Carlo simulation, while ATLAS applied a selection requiring tracks in the central region with defined $\pt$ and $\eta$ cuts, which is less Monte Carlo specific than the diffractive definition. With more data, CMS managed to extend its $\pt$ reach for the $\pt$ distribution to $\sim100\GeV/c$ using jet triggers\cite{cms:pt}. Multiplicity distributions, as well as the $\langle \pt \rangle$ evolution with the multiplicity were also published\cite{atlas:dndeta,cms:nch}, each experiment with their own event selection, and again for $\sqrt{s}=0.9, 2.36$, and $7\TeV$. It was finally decided by the Minimum Bias Underlying Event Working Group\cite{other:mbuewg} (MBUEWG) to have a common event selection to compare between experiments, ALICE included: events with at least one track in $|\eta|<0.8$ with $\pt>0.5\GeV$, which led to CMS publishing new results\cite{cms:dndeta_1tr}. Figure\ref{fig:mb} (Top Right) presents such comparison for the $\eta$ distribution: ATLAS, ALICE and CMS agree very well.

Although data at $0.9\TeV$ is compatible with previous experiments like UA5\cite{other:dndeta_ua5}, the Monte Carlo generators do not describe properly the results, this whatever the event selection. The increase in particle production with energy is underestimated by all generator tunes, which is clearly visible in the $7\TeV$ $\eta$ density results, as well as in the multiplicity distributions where high multiplicity tails in data are not reproduced by any Monte Carlo, as shown in Fig.\ref{fig:mb} (Bottom Left). Neither the $\pt$ distribution nor its evolution with multiplicity is well predicted by the Monte Carlo generators: pre-LHC tunes tend to predict too many high $\pt$ particles, particularly at high multiplicity, as seen in Fig.\ref{fig:mb} (Top Right), except for \PHOJET. As big effort on generator tuning was performed by both experiments, new tunes describe much better the $7\TeV$ data, but none are able to do so perfectly for all distributions, event selections and energies.

In order to look at the effect of diffraction, ATLAS analyzed two samples: a diffraction suppressed one requiring at least six tracks in the event, which decreases the fraction of diffractive events remaining (although depending on the diffraction model used in Monte Carlo generators, a substantial part of diffractive events have a multiplicity higher than six); and a diffraction enriched sample\cite{atlas:diffraction}, results being not corrected for detector effects. The diffraction suppressed multiplicity distribution from Fig.\ref{fig:mb} (Bottom Left) indicates that though new tunes are much better, their lack of particle production is blatant, and Monte Carlo diffraction models can not be the only source of it. The results using diffraction enriched samples exhibit the alikeness of $\PYTHIAeight$ and $\PHOJET$ diffraction models, which are much closer to data than the $\PYTHIAsix$ one, as illustrated by the $\pt$ distribution in Fig.\ref{fig:mb} (Bottom Right).

\begin{figure}[Htbp]
\center{
\resizebox{1\columnwidth}{!}{
  \subfloat{\includegraphics{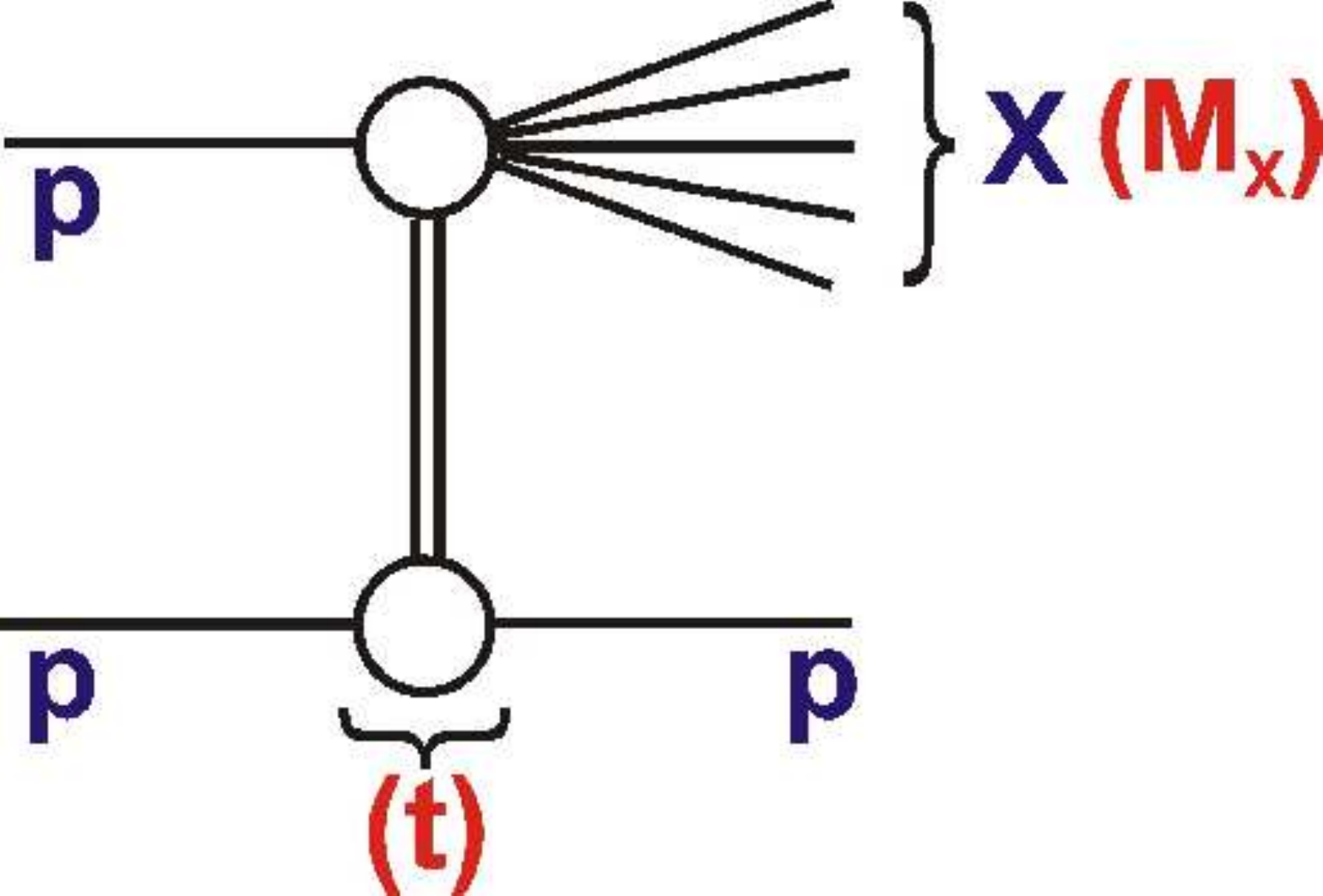}}
  \subfloat{\includegraphics{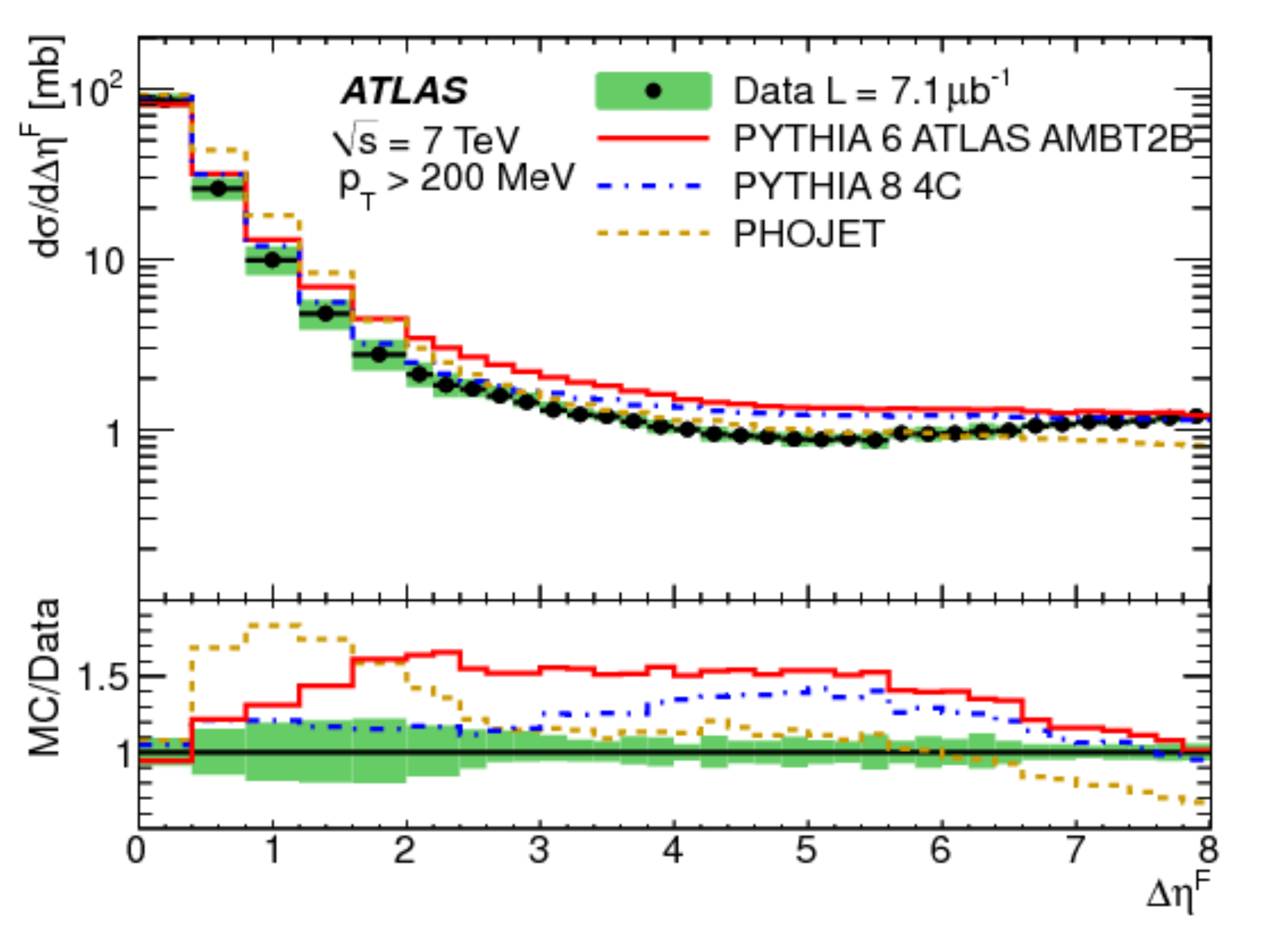}} }
}
\caption{\scriptsize{\bf(Left)} Schematic view of a single diffractive event; {\bf(Right)} Evolution of the cross-section with the pseudorapidity gap for $\sqrt{s}=7\TeV$.}
\label{fig:mb2}       
\end{figure}

ATLAS and CMS also measured the total inelastic cross-section. To compute this cross-section, ATLAS first looked at events with $\xi>5\cdot 10^{-6}$, where $\xi=M_{x}^2/s$ and $M_x$ represents the higher mass of the two hadron systems separated using the largest rapidity gap in the event, as explained in Fig.\ref{fig:mb2}(Left), and then extrapolated to the total acceptance\cite{atlas:xsec_1}. CMS used the assumption that pile-up events are randomly distributed according to a Poisson distribution to compute the cross-section for events with at least two tracks of $\pt>200\MeV$ and $|\eta|<2.4$, and then extrapolated the total inelastic cross-section using a model-dependent method\cite{cms:xsec}. The total inelastic cross-section was found to be {69.1~$\pm~2.4$~(Exp.)~$\pm~6.9$~(Ext.)~$mb$} for CMS, and {68~$\pm~2.0$~(Syst)~$\pm~2.4$~(Lum)~$\pm~4$~(Ext.)~$mb$} for ATLAS, both being overestimated by most Monte Carlo models. ATLAS also studied diffractive events by measuring the cross-section with respect to the pseudorapidity gap $\deta^F$ in the event\cite{atlas:xsec_2}. Monte Carlo generators appear to show quite different behaviors, while not describing correctly the data over the full $\deta^F$ range, as exhibited in Fig.\ref{fig:mb2}(Right).

Strange particle production has been studied in detail by ATLAS\cite{atlas:strange} and CMS\cite{cms:strange}. $K^{0}_{s}$, $\Lambda$ and $\Xi^{-}$ yields show good agreement with Particle Data Group values. Their rapidity, $\pt$ and multiplicity distributions have been measured with different event selection, and while presenting good agreement with other experiments, Monte Carlo tunes have difficulties to reproduce them, predicting too few strange particles and harder $\pt$ spectra, the differences with data increasing with the mass of the strange particle. Ratios of rapidity density between  $\Lambda$ and $K^{0}_{s}$, and $\Xi^{-}$ and $\Lambda$ for both $0.9\TeV$ and $7\TeV$ seems to indicate an energy-independent production of strange particles, which is in contradiction with Quark-Gluon Plasma (QGP) production expectancies. $\overline{\Lambda}/\Lambda$ ratios are consistent with unity in data, indicating that no significant transport of baryon number to mid-rapidities is present, in accordance with standard model  predictions and measurements from other experiments.

\begin{figure}[Htbp]
\center{
\resizebox{1\columnwidth}{!}{
  \subfloat{\includegraphics{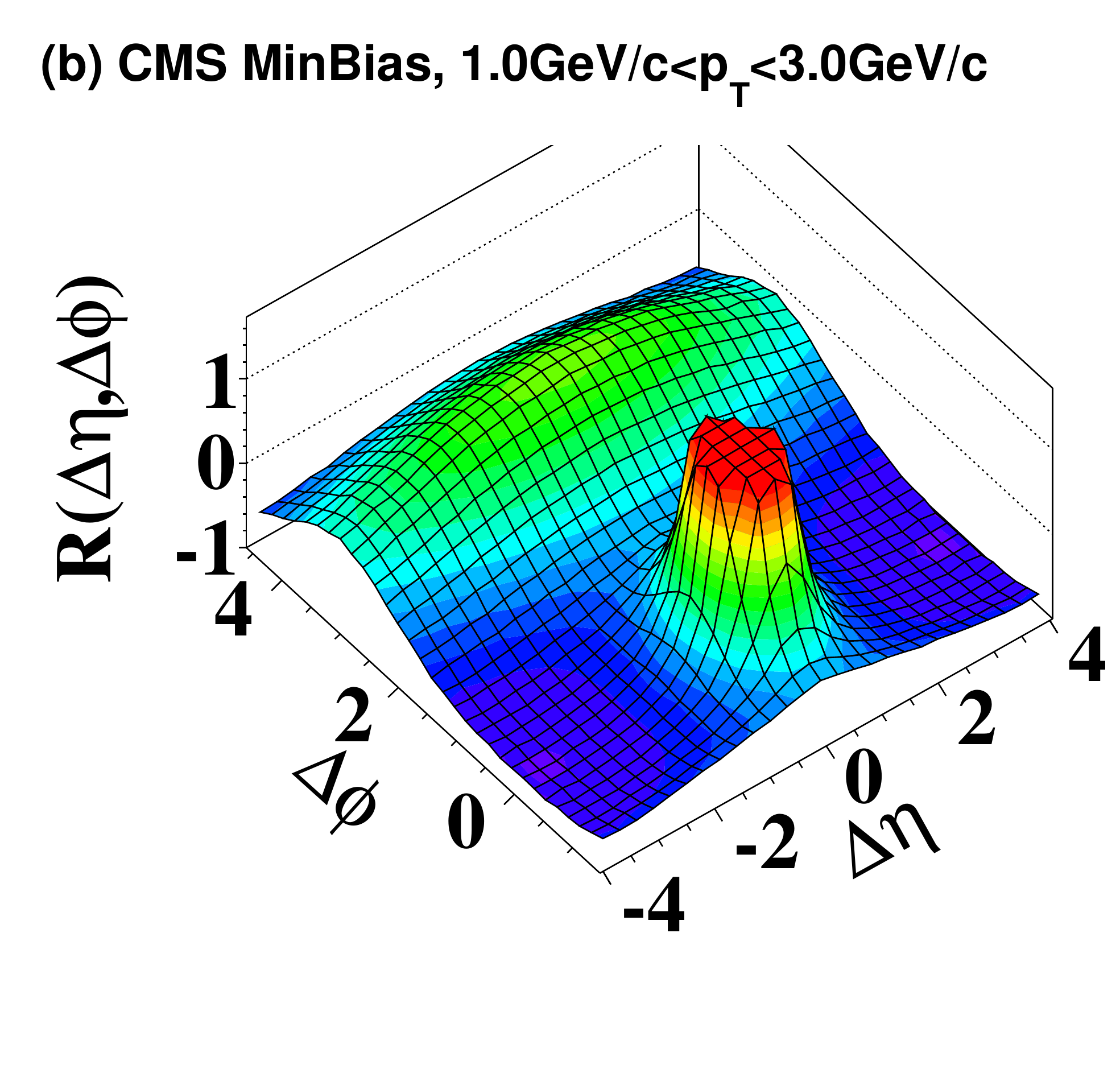}}
  \subfloat{\includegraphics{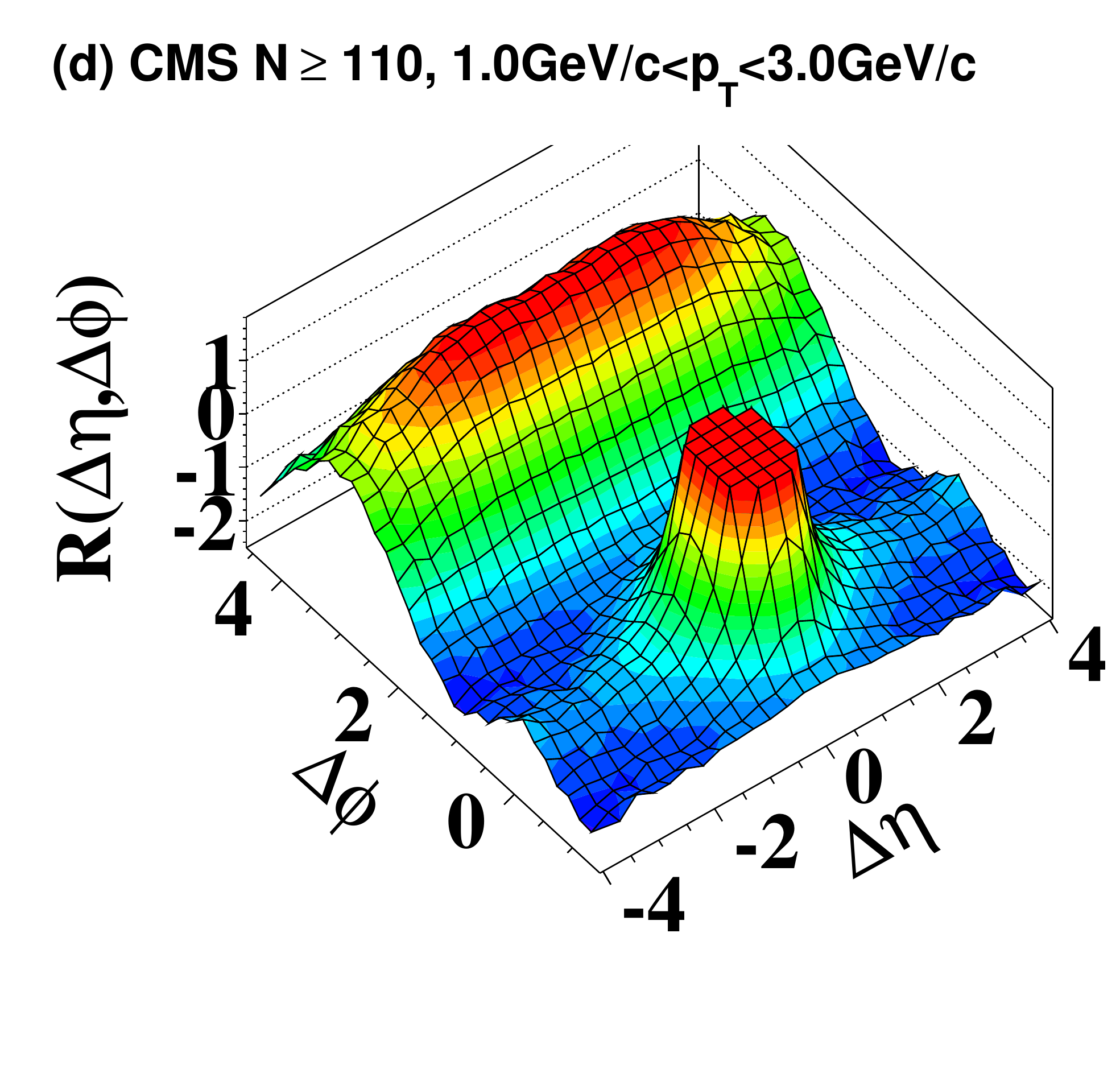}} }
}
\caption{\scriptsize Two-particle correlation in function of the $\deta$ and $\dphi$ of the pair at intermediate $\pt$ range $1<\pt<3\GeV$, for the whole minimum bias sample {\bf(Left)} and for high multiplicity events only {\bf(Right)}.}
\label{fig:mb3}       
\end{figure}

One of the first unexpected result from the LHC era appeared in the two-particle correlation study. Using pair of particles from the same event as signal and from different events as background, ATLAS\cite{atlas:corr} and CMS\cite{cms:corr} constructed a two-particle correlation function, and measured its value for different $\deta$ and $\dphi$ angular separation of the two particles, as illustrated in Fig.\ref{fig:mb3}(Left). It showed the expected jet structure for $\deta=\dphi=0$, as well as the away structure for $\dphi=\pi$. Though Monte Carlo generators do present the jet and away structure, the strength of the correlation is not well reproduced. Moreover, when looking at particles in a precise range of intermediate $\pt$ and high multiplicity ($1<\pt<3\GeV$ and $N_{ch}>110$), a clear ridge-like structure emerges at $\dphi=0$ and $\deta>\pi/2$, as shown in Fig.\ref{fig:mb3}(Right) by CMS (ATLAS did not have enough statistics in this narrow phase-space), which is absolutely not reproduced by any Monte Carlo generators simulating $\pp$ collisions. Although this feature is well known in heavy ion collisions, it was the first time it was observed in $\pp$ collisions. Many explications were since then proposed, but at the time none has rise above all else and make consensus.

\subsection{Underlying Event results}\label{ue}

The Underlying Event (UE) is usually defined, in presence of a hard probe in the event, as everything but this probe. Depending on the subdetector used, different hard probes can be considered to get the scale of the event: ATLAS used the highest $\pt$ track\cite{atlas:ue_track} or calorimeter cluster\cite{atlas:ue_calo}, while CMS preferred looking at trackjets\cite{cms:ue_trackjet} or muon pairs in Drell-Yan events\cite{cms:ue_drellyan}. Each of those choices have different scales, and thus can not be directly compared: since trackjets are clusters of tracks, they reach higher $\pt$ values than single tracks; calorimeter clusters also take into account neutral particles undetected in the tracker; Drell-Yan events have very clean signals and use the muon systems to get the event scale. Three topological regions  of $120\de$ are then defined with respect to the hard probe: the towards region ($0\de<|\dphi|<60\de$) where the hard part of the event is, the away region  ($120\de<|\dphi|<180\de$) balancing the towards $\pt$, and the transverse region ($60\de<|\dphi|<120\de$) where the Underlying Event lies, $\dphi$ being the difference in azimuthal angle $\phi$ between the hard probe and the track/cluster. For Drell-Yan events, both transverse and towards regions are sensitive to UE.

\begin{figure}[Htbp]
\resizebox{1\columnwidth}{!}{
  \subfloat{\includegraphics{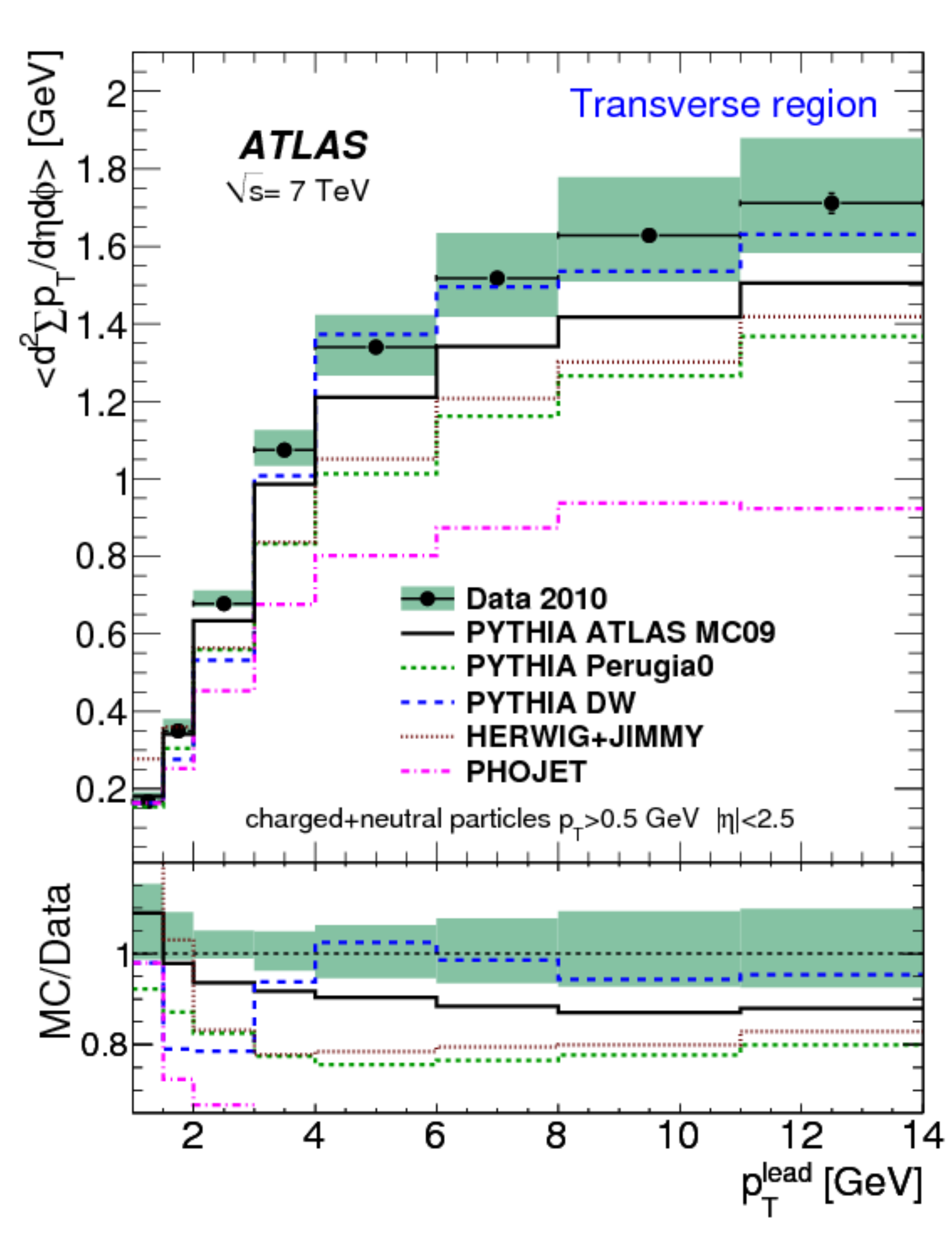}}
  \subfloat{\includegraphics{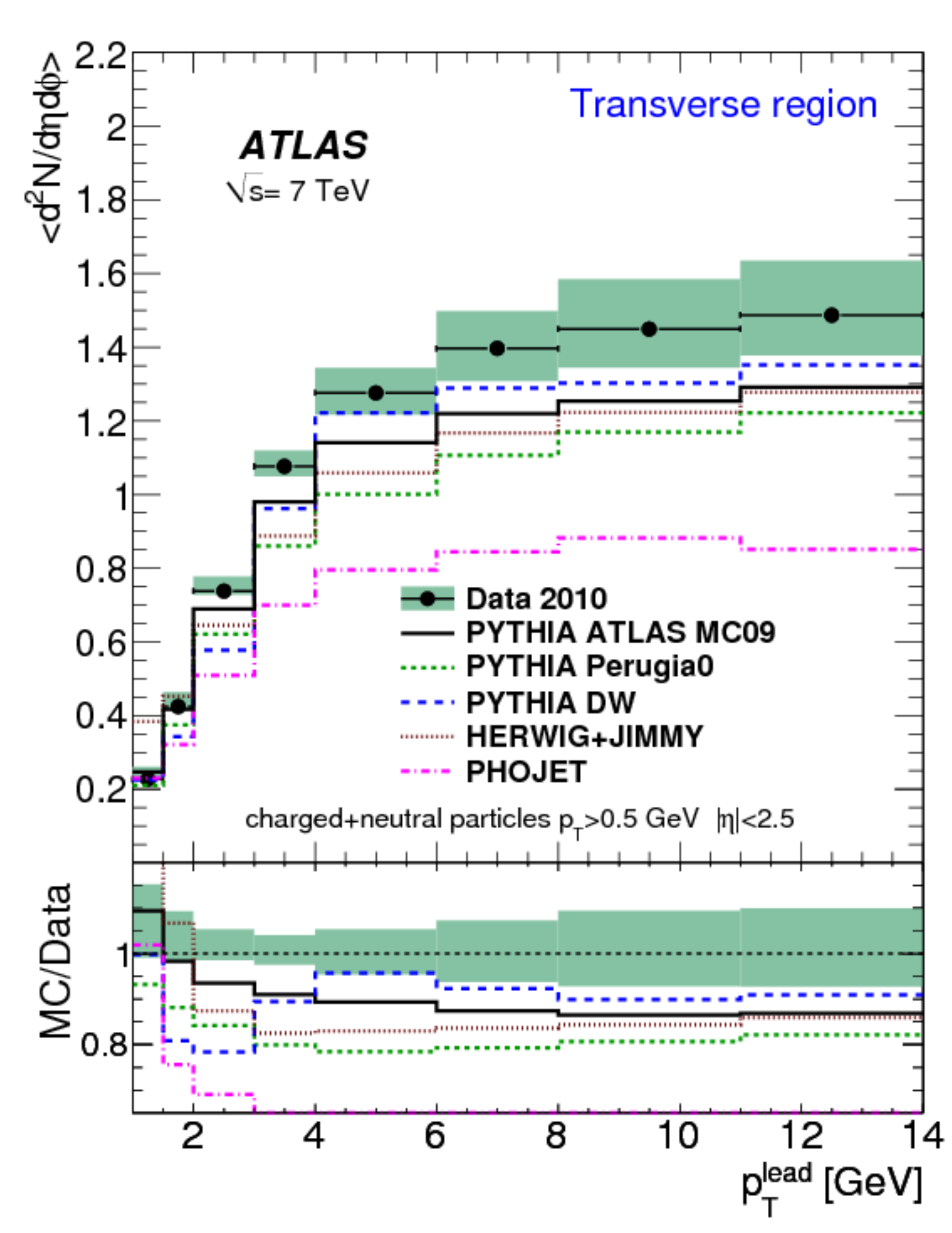}} }
\resizebox{1\columnwidth}{!}{
  \subfloat{\includegraphics{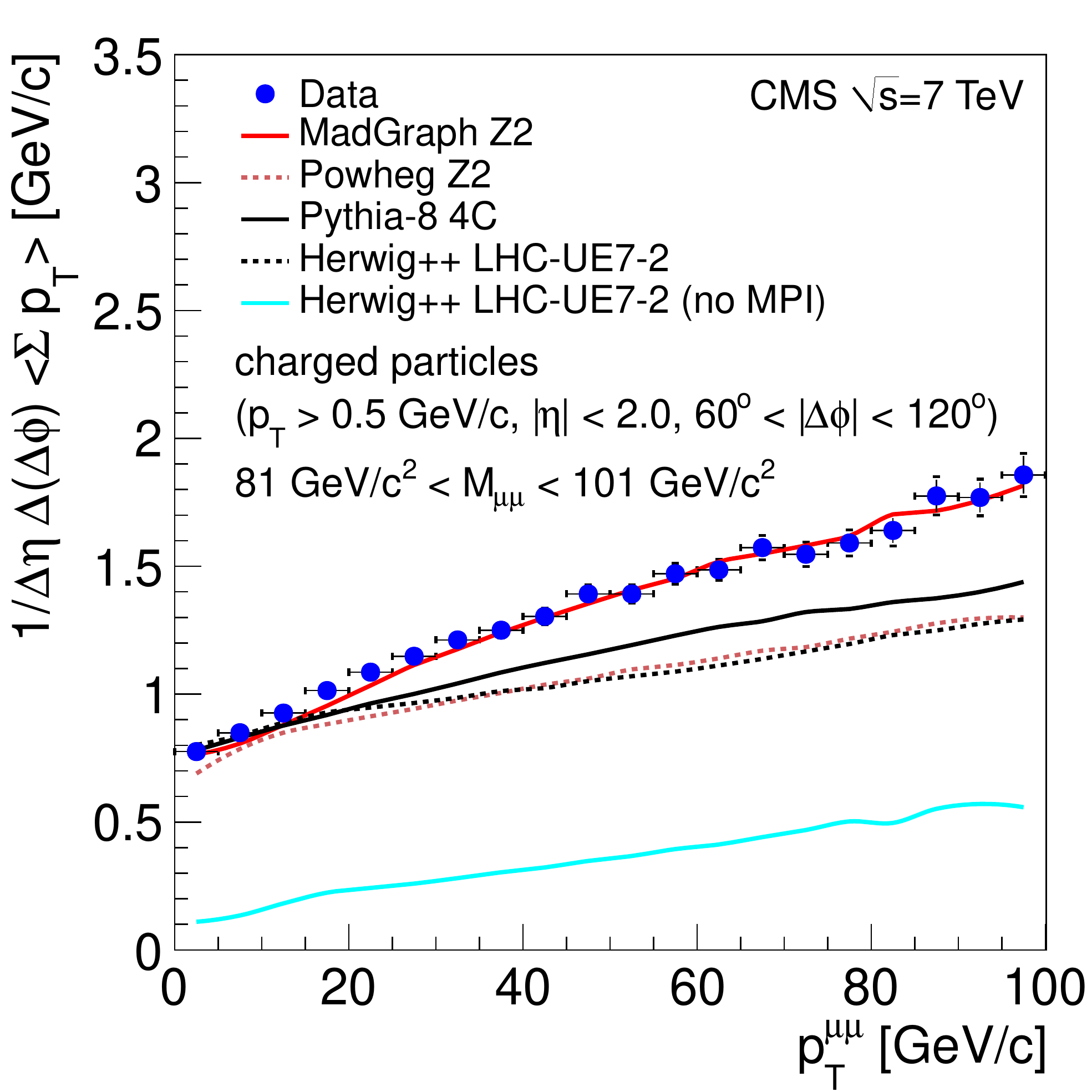}}
  \subfloat{\includegraphics{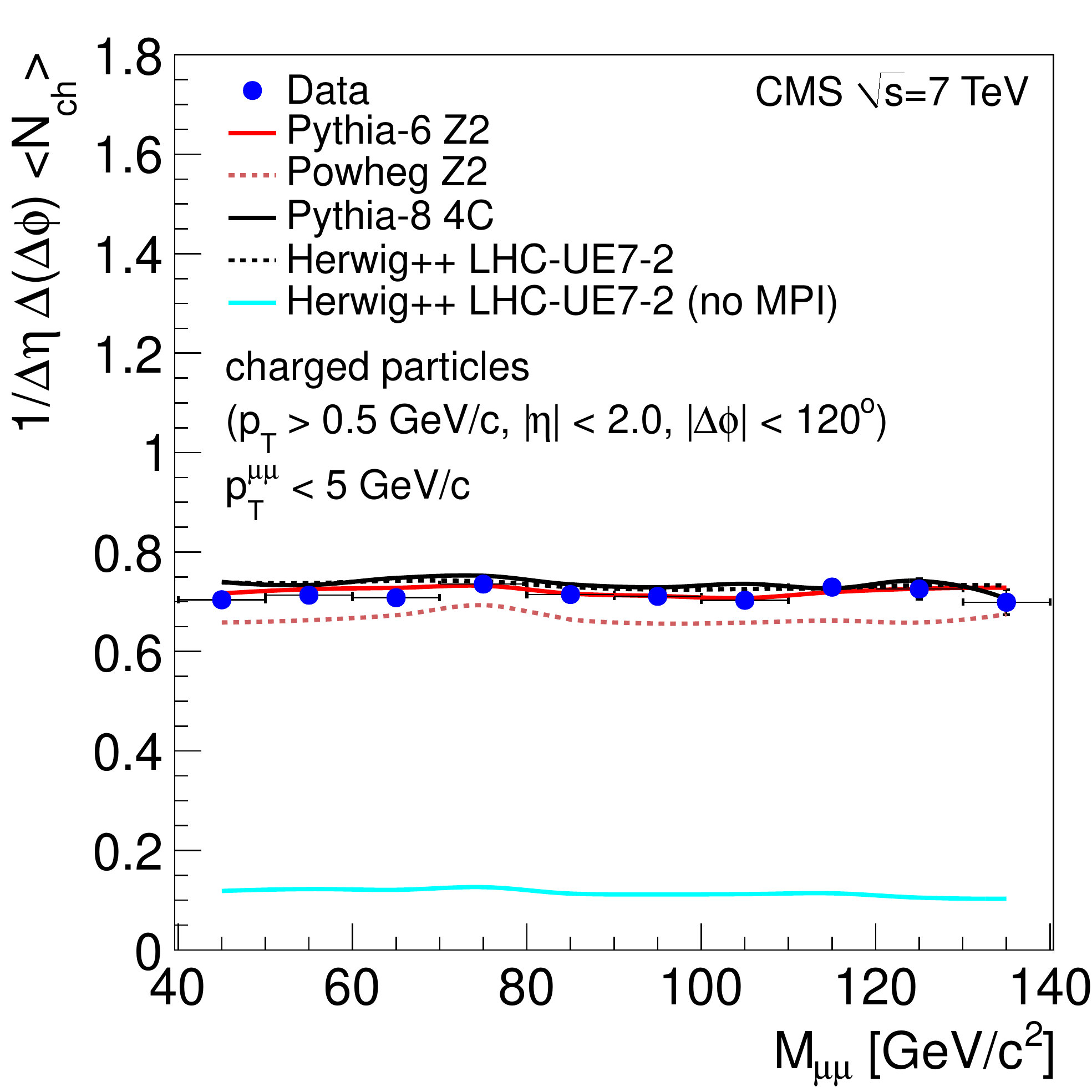}} }
\caption{\scriptsize{\bf(Top)} $\langle\sum\pt\rangle$ and $\langle N_{ch}\rangle$ densities with respect to $\pt$ of leading calorimeter cluster for the transverse region; {\bf(Bottom)} $\langle\sum\pt\rangle$ w.r.t. $\pt$ of muon pair and $\langle N_{ch}\rangle$ w.r.t. mass of muon pair, for Drell-Yan events.}
\label{fig:ue}       
\end{figure}

The Monte Carlo generators seem to underestimate the particle density, especially in the transverse region. When looking, in the transverse region, at the evolution with respect to the leading $\pt$ probe of the mean multiplicity and mean $\pt$ sum, the fast rise at low $\pt$ scale is attributed to multiple partonic interaction (MPI), before reaching a plateau, as Fig.\ref{fig:ue} (Top) illustrates in case of a calorimeter cluster probe. Moreover, once again no generator tune describes properly neither the rise nor the plateau. For Drell-Yan events, shown in Fig.\ref{fig:ue} (Bottom), the increase with the $\pt$ of the muon pair comes from ISR increase, while $\langle N_{ch}\rangle$ and $\langle\sum\pt\rangle$ densities being independent of the muon pair mass is explained by the already saturated MPI in the selected event scale.

\section{Conclusion}\label{conclusion}

ATLAS and CMS have proven their general purposefulness with many publications of soft QCD physic results. With several event selections, minimum bias distributions of $\eta$, $\pt$, $N_{ch}$ and $\langle\pt\rangle VS N_{ch}$  at $\sqrt{s}=0.9, 2.36,$ and $7\TeV$ were compared with other experiments when possible, as well as a multitude of Monte Carlo generator tunes, none of which are able to describe the softer $\pt$ and huge rise of particle production with energy. This led to a community-wide effort to improve and tune the available Monte Carlo generators. Extrapolated inelastic cross-section were measured and compatible, though overestimated by Monte Carlo models. Diffraction was also studied, with minimum bias observables for either diffraction enriched or suppressed data samples, as well as the rapidity gap cross-section. Due to excellent tracker performances, strange particle production was examined, Monte Carlo discrepancies with data increasing with the particle mass. An unexpected ridge-like structure was found during the two-particle correlation analysis for mid-$\pt$ and high multiplicity particles, and albeit attracting many attention from theory, its origin is still being discussed. Finally, underlying events have been studied with different hard probes, and despite Monte Carlo generators being tuned with those results, none are able to describe perfectly the data.

\end{document}